\documentclass[12pt,a4paper]{article}
\usepackage{amsmath,amssymb,amsthm}
\usepackage[margin=1.0in]{geometry}
\usepackage{cite}
\usepackage{graphicx}
\usepackage{enumerate}
\allowdisplaybreaks
\usepackage[colorlinks=true
,urlcolor=blue
,anchorcolor=blue
,citecolor=blue
,filecolor=blue
,linkcolor=blue
,menucolor=blue
,pagecolor=blue
,linktocpage=true
,pdfproducer=medialab
,pdfa=true
]{hyperref}
\numberwithin{equation}{section}

\def\a{\alpha}
\def\b{\beta}
\def\g{\gamma}
\def\d{\delta}
\def\e{\epsilon}

\def\th{\theta}

\def\k{\kappa}
\def\l{\lambda}
\def\m{\mu}
\def\n{\nu}

\def\r{\rho}

\def\s{\sigma}
\def\t{\tau}

\def\ph{\phi}

\def\Th{\Theta}

\def\S{\Sigma}

\def\L{\Lambda}

\def\be{\begin{equation}}
\def\ee{\end{equation}}
\def\bea{\begin{eqnarray}}
\def\eea{\end{eqnarray}}

\def\pa{\partial}

\def\lp{\left(}
\def\rp{\right)}
\def\ls{\left[}
\def\rs{\right]}
\def\nn{\nonumber}
\def\ie{{\it i.e., }}

\makeatletter
\renewcommand\section{\@startsection {section}{1}{\z@}%
	{-3.5ex \@plus -1ex \@minus -.2ex}
	{2.3ex \@plus.2ex}%
	{\normalfont\large\bfseries}}
\renewcommand\subsection{\@startsection{subsection}{2}{\z@}%
	{-3.25ex\@plus -1ex \@minus -.2ex}%
	{1.5ex \@plus .2ex}%
	{\normalfont\bfseries}}
\makeatother

\begin{document}

\begin{center}
\addtolength{\baselineskip}{.5mm}
\thispagestyle{empty}
\begin{flushright}
\end{flushright}

\vspace{20mm}

{\Large  \bf Higher dimensional charged static and rotating solutions in mimetic gravity}
\\[15mm]
{Hamid R. Bakhtiarizadeh\footnote{h.bakhtiarizadeh@kgut.ac.ir}}
\\[5mm]
{\it Department of Nanotechnology, Graduate University of Advanced Technology,\\ Kerman, Iran}

\vspace{20mm}

{\bf  Abstract}
\end{center}

We find new solutions to the Einstein-Maxwell equations in the presence of mimetic field in $ D $ dimensions, all of which are asymptotically anti–de Sitter. We derive the solutions in five-dimensional spacetime, in detail. By extending the calculations to six and seven dimensions, we obtain a general form for solutions in dimensions larger than four. The results describe electrically charged static and rotating solutions, which have spherical, toroidal or cylindrical horizons. The solutions, depending on their global identifications, can be regarded as black holes, or black strings/branes. Some physical properties of solutions such as horizons, singularities as well as entropy, mass, and angular momenta of rotating solutions are also investigated.

\vfill
\newpage


\section{Introduction}\label{int}

One of the most interesting theories of gravity, which without introducing any additional matter field describe the dark components of the universe as a geometrical effect, is mimetic gravity \cite{Chamseddine:2013kea}. In this theory, the conformal degree of freedom of gravity is isolated by introducing the following relation between physical metric $ g_{\m \n} $, auxiliary metric $ \tilde{g}_{\m \n} $, and a scalar field known as mimetic field,
\be
g_{\m \n}=\pm \tilde{g}_{\m \n}\tilde{g}^{\a \b}\pa_{\a}\ph\pa_{\b}\ph.
\ee
As a consistency condition, the above equation implies that the mimetic field must satisfy the following constraint,
\be
g^{\m \n}\pa_{\m}\ph\pa_{\n}\ph=\pm 1. \label{const}
\ee
The equations of motion of the gravitational action are equivalent to those one can obtain from the action in terms of the physical metric with the imposition of constraint (\ref{const}), by using a Lagrange multiplier \cite{Chamseddine:2013kea}. The conformal degree of freedom, even in the absence of matter, is dynamical and this mimics the phenomenon of cold dark matter evolution of the universe in the background. It has been also shown that the scalar field can mimic the gravitational behavior of any form of matter \cite{Chamseddine:2014vna,Lim:2010yk}. 

In the following, we briefly review some attempts in obtaining the gravitational and cosmological solutions in the context of mimetic model. Spherically symmetric static solutions with various topologies in the framework of mimetic gravity are analyzed in \cite{Myrzakulov:2015sea,Myrzakulov:2015kda,Deruelle:2014zza,Nashed:2018qag,Nashed:2018urj,Nashed:2018aai}. The vacuum static and spherically symmetric solutions in the mimetic gravity scenario are explored in \cite{Gorji:2019rlm}, based on the conformal invariance principle. Modified gravity models including mimetic scalar potential are also studied for Gauss-Bonnet \cite{Astashenok:2015haa}, $ f(R) $  \cite{Odintsov:2015cwa,Nojiri:2017ygt,Odintsov:2018ggm,Oikonomou:2016fxb}, Born-Infeld \cite{Chen:2017ify} and Ho\v{r}ava \cite{Chamseddine:2019gjh} theories. It also is shown that, in the presence of mimetic field, a schwarzschild black hole can be exist without singularity \cite{Chamseddine:2016ktu}. Recently, in \cite{Sheykhi:2019gvk,Sheykhi:2020fqf,Sheykhi:2020dkm} the authors obtained topological black hole, black string as well as there-dimensional BTZ-like solutions in the framework of mimetic gravity. From the cosmological point of view, the authors indicate that the singularities in cosmological solutions can be easily removed by considering the longitudinal degree of freedom of gravity to be dynamical \cite{Chamseddine:2016uef}. In \cite{Dutta:2017fjw}, the dynamical behavior of mimetic gravity with a general form of potential for the mimetic scalar field has been investigated. It also has been observed that, in the presence of some classified higher-derivative corrections, the mimetic cosmological perturbations can be stable \cite{Gorji:2017cai}. Different aspects of mimetic cosmology are also explored in \cite{Abbassi:2018ywq,Sadeghnezhad:2017hmr,Zhong:2018tqn,Matsumoto:2016rsa,Gorji:2018okn,Gorji:2019ttx}. For a comprehensive survey on mimetic gravity and its applications to cosmology and astrophysics, the reader can refer to \cite{Sebastiani:2016ras}.

Most efforts in finding a quantum theory of gravity have been directed at studying theories in which the dimension of spacetime is larger than four. It also appears that black holes will play a crucial role in understanding nonperturbative effects in a quantum theory of gravity. Thus, exploring black holes in $ D>4 $ dimensions is important to obtain a full understanding of these theories \cite{Emparan:2008eg}. On the other side, the large black holes that have finite-area horizons in the supergravity approximation only arise for $ D \leq 5 $. This is one reason why there has been a lot of interest for studying the black ring case. Another reason is that nonspherical horizon topologies become possible for $ D > 4 $. Furthermore, black branes, which are generalizations of black holes to higher dimensions, play a crucial role in the context of the AdS/CFT conjecture \cite{Becker:2006dvp}. However, motivations for studying higher dimensional general relativity also originate from this theory itself since it exhibits much more richer dynamics than in the four-dimensional case. For example, the existence of black rings or other black objects with nonspherical horizon topologies leading to the violation of the four-dimensional black hole uniqueness theorem in higher dimensions. Brane-world scenarios also consider that our four-dimensional universe lies on a brane embedded in a higher dimensional spacetime \cite{Malek:2012jv}.

The generalization of black hole solutions to higher dimensions first described by Tangherlini \cite{Tangherlini:1963bw} for static vacuum black holes, and Myers and Perry \cite{Myers:1986un} for stationary vacuum charged static black holes. With the advent of brane-world theories, black hole solutions in higher dimensions received much interest \cite{Dimopoulos:2001hw,Giddings:2001bu}. Models of modified gravity with extra dimensions appear when considering string theory. Their black holes solutions are also known for some low energy effective actions related to string theory \cite{Cvetic:1996xz,Breckenridge:1996sn,Breckenridge:1996is,Kallosh:1996vy,Klemm:2000vn,Cvetic:2004hs,Cvetic:2004ny} and are interesting in the context of gauge/gravity duality because yield to a quantitative evidence for $ {\rm AdS}_{D}/{\rm CFT}_{(D-1)} $ \cite{Maldacena:1997re}. 

In the present work, we are going to find the charged static and rotating solutions in the context of mimetic gravity in spacetimes with dimensions larger than four and investigate their physical properties. The solutions are indeed the generalizations of those obtained in \cite{Sheykhi:2020fqf,Sheykhi:2020dkm} to higher dimensions. Depending on their identifications, one can interpret them as black holes, or black strings/branes. The structure of paper is arranged as follows. First, in Sec. \ref{feq}, we introduce the mimetic Einstein-Maxwell action as well as its field equations in $ D $ dimensions. Then, in Sec. \ref{solin5}, we will derive the solutions for five-dimensional spherical as well as cylindrical symmetric spacetimes. In Sec. \ref{ext}, we extend our solutions to include higher-dimensional spacetimes. Finally, we summarize closing remarks in section \ref{con}.

\section{Action and field equations}\label{feq} 

The action of Einstein-Maxwell (EM) gravity in the presence of mimetic field for $ D $-dimensional asymptotically anti–de Sitter (AdS) spacetimes is given by\footnote{We work in units in which $ M_P=1/\sqrt{8\pi G_D}=1 $ and choose the metric signature as $ (-,+,\cdots,+) $.} 
\be
I_D= \int d^D x \sqrt{-g}\ls R+\frac{(D-1)(D-2)}{l^2}-F_{\m \n}F^{\m \n}+\l(g^{\m \n}\pa_{\m} \ph \pa_{\n} \phi-\e)\rs,\label{Daction}
\ee
where $ g $ is the determinant of the physical metric $ g_{\m \n} $, $ R $ is the Ricci scalar, $ \l $ is the Lagrange multiplier and $ \L=-(D-1)(D-2)/2l^2 $ is the negative cosmological constant describing AdS spacetimes. $ F_{\m \n}=\pa_{\m} A_{\n}-\pa_{\n} A_{\m} $ is the electromagnetic field strength tensor corresponding to the gauge potential $ A_{\m} $. Here $ \e=+1,-1 $, respectively, denote to the spacelike and timelike nature of vector field $ \pa_{\m} \ph $. Variation of the above action with respect to Lagrange multiplier $ \l $ yields the following constraint on mimetic field\footnote{The symbolic calculations of this paper have been carried out using the excellent Mathematica package ``xTras" \cite{Nutma:2013zea}.}, 
\be
g^{\m \n}\pa_{\m} \ph \pa_{\n} \phi=\e.\label{lambdaeq}
\ee
This equation implies that the physical metric must satisfy the constraint (\ref{lambdaeq}) and confirms that the scalar field is not dynamical. By varying the action (\ref{Daction}) with respect to the physical metric $ g_{\m \n} $, the scalar field $ \ph $ and the gauge field $ A_{\m} $, one can get the Einstein, scalar and Maxwell field equations as
\be
G_{\m \n}=T_{\m \n}+T_{\m \n}^{\ph},\label{geq}
\ee
\be
\nabla_{\m} (\l \pa^{\m}\ph)=0,\label{phieq}
\ee
\be
\nabla_{\m}F^{\m \n}=0.\label{Feq}
\ee
 Here, $ G_{\m \n}= R_{\m \n}- R g_{\m \n}/2$, $ T_{\m \n}^{\ph}=(D-1)(D-2)g_{\m \n}/2l^2+\l \pa_{\m} \ph \pa_{\n} \phi $ and $ T_{\m \n}=2 F_{\m \g}F_{\m}{}^{\g}-g_{\m \n}F_{\a \b}F^{\a \b}/2$ are respectively the Einstein tensor, scalar field energy-momentum tensor, and Maxwell energy-momentum tensor. Taking the trace of Eq. (\ref{geq}), when one takes into account Eq. (\ref{lambdaeq}), leads to
\be
\l=\e \ls G-T-\frac{D(D-1)(D-2)}{2l^2}\rs,\label{traceeq}
\ee
where $ G=G_{\m}{}^{\m} $ and $ T=T_{\m}{}^{\m} $ respectively denote the trace of Einstein tensor and Maxwell energy momentum tensor. By inserting Eq. (\ref{traceeq}) into Eqs. (\ref{geq}) and (\ref{phieq}), we arrive at  
\be
G_{\m \n}=T_{\m \n}+\frac{(D-1)(D-2)}{2l^2} g_{\m \n}+\e\ls G-T-\frac{D(D-1)(D-2)}{2l^2}\rs\pa_{\m} \ph \pa_{\n} \phi,\label{geqfinal}
\ee
\be
\nabla_{\m} \ls\lp G-T-\frac{D(D-1)(D-2)}{2l^2}\rp \pa^{\m}\ph\rs=0.\label{phieqfinal}
\ee 
Our aim in the following is to solve the above equations in spherically and cylindrically symmetric spacetimes and for various dimensions. 

\section{Solutions in five dimensions}\label{solin5}

By making restrictions, in particular with respect to the horizon topology, one can generalize EM theory to five dimensions \cite{Morisawa:2004tc,Emparan:2001wn,Elvang:2003yy,Emparan:2004wy,Kunz:2005ei,Andre:2020czm}. In the context of mimetic gravity, one can get the final form of five-dimensional version of field equations as well as scalar field equation by setting $ D=5 $ in Eqs. (\ref{geqfinal}) and (\ref{phieqfinal}). In the following two subsections, we are going to solve the field equations for spherically and cylindrically symmetric spacetimes.   

\subsection{Charged static solutions in spherically symmetric spacetimes}

Here, we are looking for charged solutions for static black holes in spherically symmetric spacetimes. In doing so, we consider the following five-dimensional metric
\be
ds^2=-f(r)g^2(r)dt^2+\frac{1}{f(r)}dr^2+r^2 d\Omega^{2},\label{sssmetric}
\ee
where $ d\Omega^{2}=d\psi^2+\sin^2\psi(d\th^2+\sin^2\th d\varphi^2) $ is the line element of a three-sphere. Here, the topology of event horizon and spacetime are given by $ S^3 $ and $ \mathbb{R}^2\times S^3 $, respectively. In the line element (\ref{sssmetric}), we also incorporate an extra degree of freedom due to the additional mimetic field, by the function $ g(r) $. 

Using the metric (\ref{sssmetric}), the constraint (\ref{lambdaeq}) yields the following equation for the mimetic field  
\be
f\ph'^2=\e,\label{fphieq}
\ee 
where a prime denotes derivative with respect to $ r $. This differential equation has the following solution
\be
\ph(r)=\pm\int\frac{dr}{\sqrt{\lvert\e f(r)\rvert}},\label{fphieqsol}
\ee
plus a constant of integration, which we observe that disappears in the field equations and accordingly we are safe to set it to zero. We are also free to choose the positive sign of the above solution without loss of generality.

\begin{figure}[htp]
	\begin{center}
		\includegraphics[width=10cm]{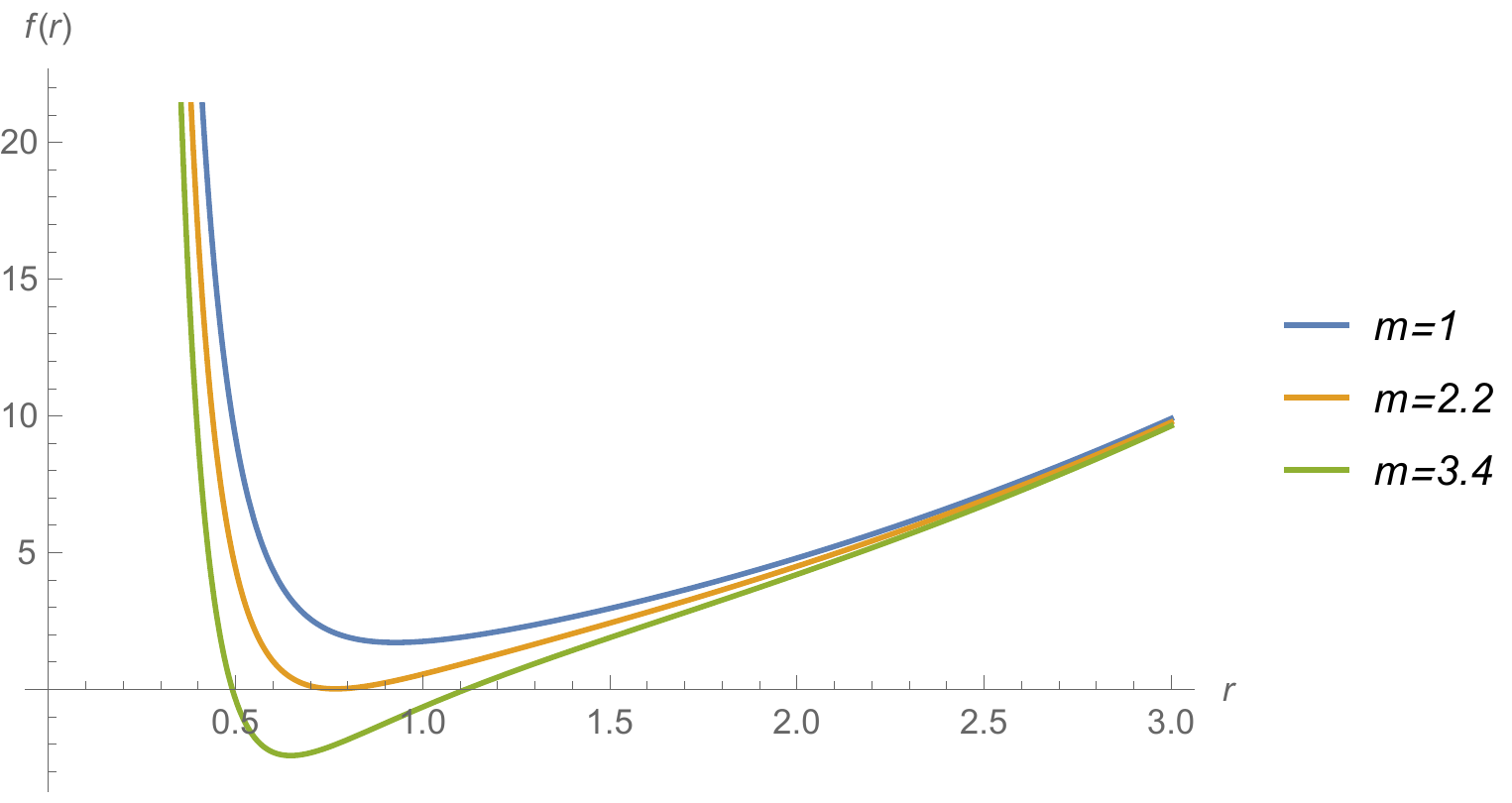}
		\caption{The diagram of function $f(r)$ for five-dimensional charged mimetic black hole with $l=1$, $ q=1.5 $, and different values of $ m $.}\label{fig1}
	\end{center}
\end{figure}

First, we consider the charged solutions. In this case, the vector potential and nonzero components of electromagnetic tensor are given by 
\bea
A_{\m}=h(r)\d_{\m}^{0};\nn\\
F_{tr}=-F_{rt}=-h'(r).\label{nonvanF}
\eea
Having had the electromagnetic field strength tensor, it is not difficult to show that the $ t $ component of Maxwell equation (\ref{Feq}) leads to the following equation
\be
r h'' g-r h' g'+3 h' g =0,
\ee
with the solution
\be
h'(r)=\frac{q}{r^3}g(r).
\ee
In obtaining the above solution we set the constant of integration equal to the electric charge $ q $ of black hole. By inserting the metric (\ref{sssmetric}), Eq. (\ref{fphieq}), the constraint $ \e^2=1 $, and electromagnetic field strength (\ref{nonvanF}) into Eq. (\ref{geqfinal}), one arrives at the following three independent equations in five dimensions
\be 
12 r^6-2q^2 l^2+l^2 r^4\lp 6-6f-3rf'\rp=0,\label{stteq}
\ee
\be
g\{48r^6 +4 l^2\lp q^2+3 r^2 \rp -3l^2r^4\ls 4 f+r\lp 5 f'+r f''\rp\rs\} -3l^2 r^5\ls\lp 4 f+3 r f'\rp g'+2rfg''\rs=0,\label{srreq}
\ee
\be
-2 l^2 r^4 g-12r^6 g-2 q^2 l^2 g+l^2 r^5 f'\lp 4g+3rg'\rp+l^2 r^6 f'' g+2l^2r^4f\ls g+r\lp 2g'+rg''\rp\rs=0.\label{spspsththphpheq}
\ee
By solving Eq. (\ref{stteq}), one finds
\be
f(r)=1+\frac{r^2}{l^2}-\frac{m}{r^2}+\frac{q^2}{3r^4}.\label{chsssfsol}
\ee
Here, $ m $ is an integration constant which is indeed the black hole mass. Plugging the solution (\ref{chsssfsol}) for metric function $ f(r) $ into Eq. (\ref{srreq}) or (\ref{spspsththphpheq}) yields
\be
\ls l^2 \lp 4 q^2-3 m r^2-6 r^4\rp-15 r^6\rs g'-\{ r l^2 \ls q^2+3 r^2 \lp r^2-m\rp\rs+3 r^7\} g''=0.
\ee
Solving this equation leads, 
\be
g(r)=1+c_1\int \frac{r^4 dr}{\lvert3r^6+l^2\ls q^2+3r^2\lp r^2-m\rp\rs\rvert^{3/2}},\label{sgsol}
\ee
where the constant of integration $ c_1 $ describes the effect of mimetic field in the solution. We also check that the above solutions satisfy Eq. (\ref{phieqfinal}) for mimetic field, as was expected. Setting $ c_1=0 $, one can easily recover the 5-dimensional charged black holes of Einstein gravity in spherically symmetric spacetimes. The horizons are the real roots of equation $ f(r_h)=0 $. Here, we find out that there are at most two horizons, corresponding a black hole with inner and outer horizon. Note that, depending on choosing the parameters, one may also get the solutions which describe extremal black hole with one horizon, or a naked singularity without an event horizon, as illustrated in Fig. \ref{fig1}. 

\begin{figure}[htp]
	\begin{center}
		\includegraphics[width=10cm]{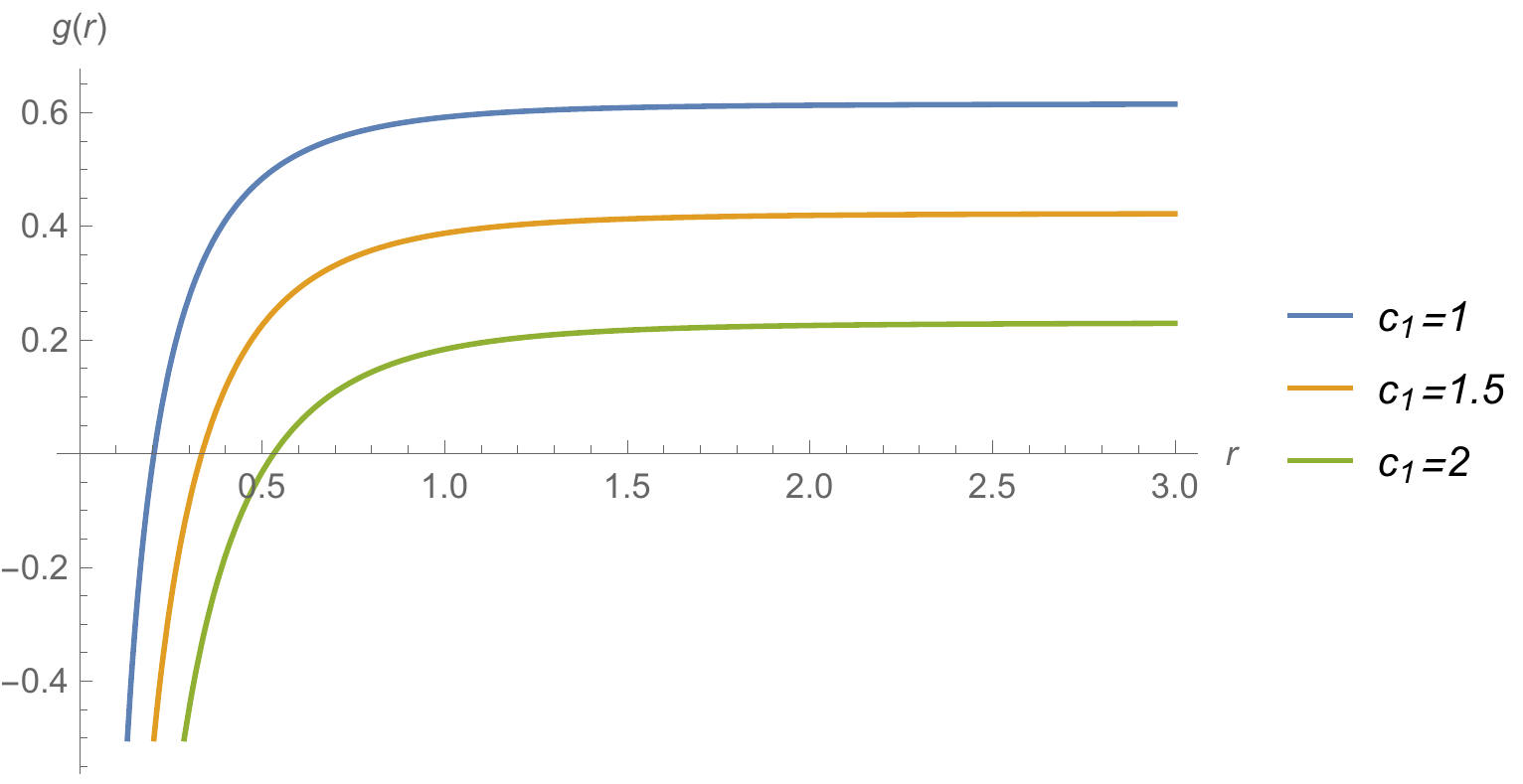}
		\caption{The diagram of function $g(r)$ for five-dimensional charged mimetic black hole at large $ r $, and for different values of $ c_1 $.}\label{fig2}
	\end{center}
\end{figure}

The integrand in Eq. (\ref{sgsol}) diverges at $ r=r_h $, while both Kretschmann and Ricci scalars have finite values at $ r=r_h $. This indicates that $ r=r_h $ is a coordinate singularity, not a curvature one. To obtain an analytical solution, we examine the asymptotic behavior of function $ g(r) $. First, we consider the large $ r $ limit for which we can approximate the absolute value term in the integrand of Eq. (\ref{sgsol}) with $ 3(r^6+l^2 r^4) $, and rewrite the function $ g(r) $ as   
\bea
g(r)&\approx& 1-c_1\frac{2 r^2+l^2}{3l^4\sqrt{3(r^4+l^2 r^2)}}\nn\\&\approx&1-\frac{2 c_1}{3 \sqrt{3} l^4}-\frac{c_1}{12 \sqrt{3} r^4}+\mathcal{O}\lp r^{-6} \rp.
\eea
Therefore, $ g(r) $ tends to the constant value $ 1-\frac{2 c_1}{3 \sqrt{3} l^4} $ as $ r \rightarrow \infty $, as can be seen from Fig. \ref{fig2}.

\begin{figure}[htp]
	\begin{center}
		\includegraphics[width=10cm]{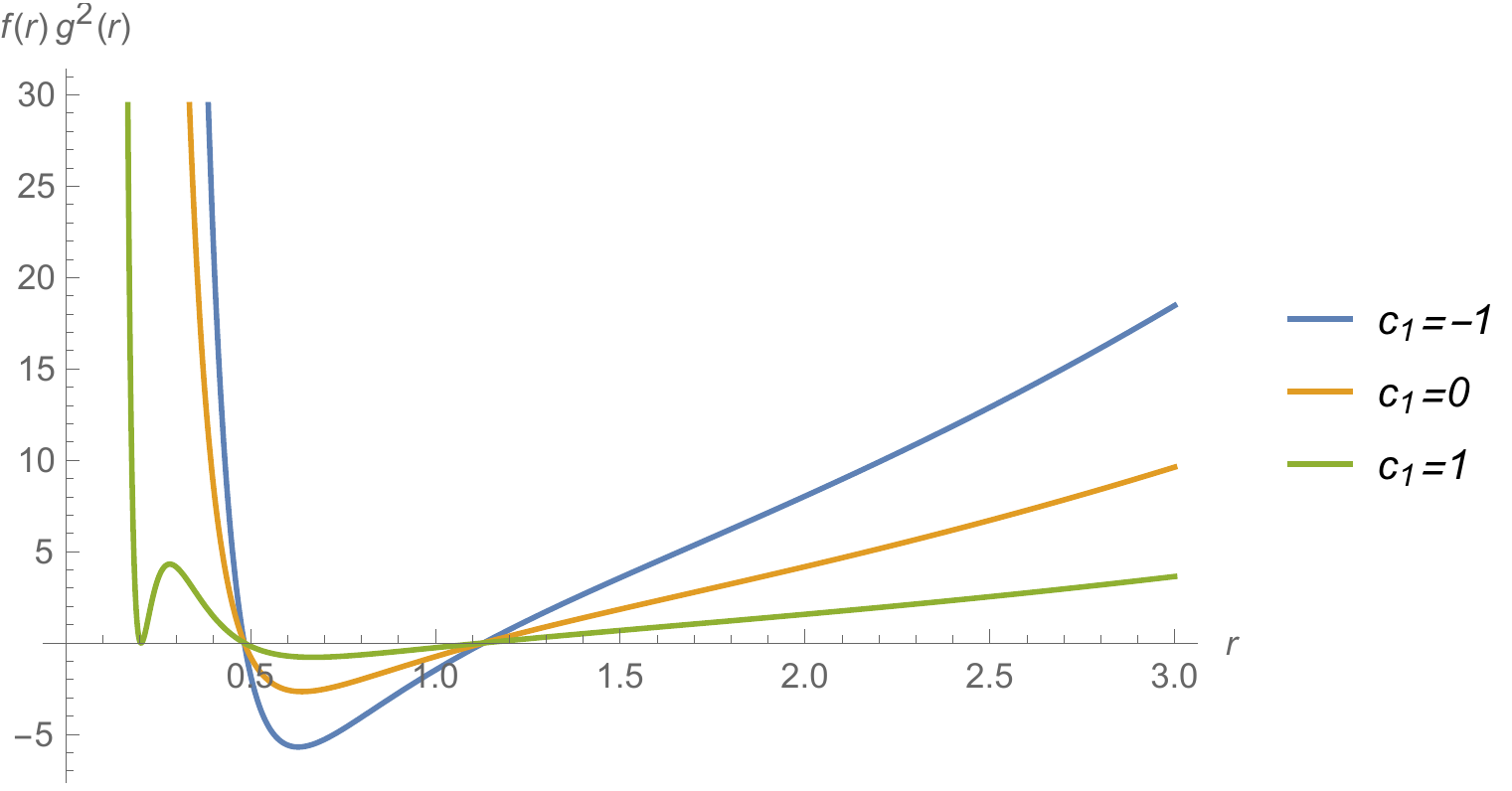}
		\caption{The behavior of function $f(r)g^2(r)$ for five-dimensional charged mimetic black hole at large $ r $ and different values of $ c_1 $. Here we choose $ m=3.5,l=1,q=1.5 $.}\label{fig3}
	\end{center}
\end{figure}

\begin{figure}[htp]
	\begin{center}
		\includegraphics[width=10cm]{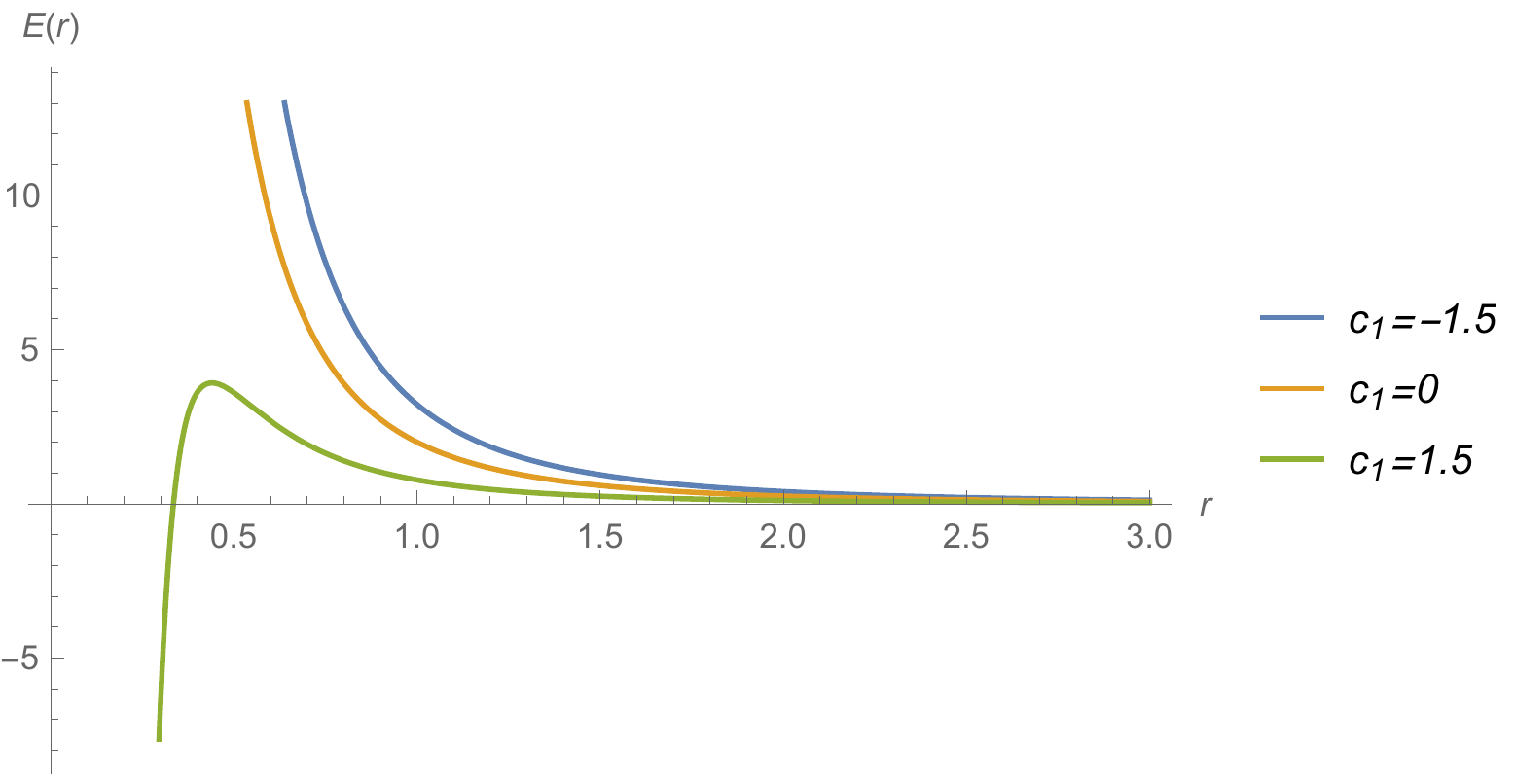}
		\caption{The behavior of Electric field  $E(r)$ for five-dimensional charged mimetic black hole at large $ r $ with $ q=1.5 $.}\label{fig4}
	\end{center}
\end{figure}

The metric function $ -g_{tt}=f(r)g^2(r) $ also takes the following form at large $ r $
\bea
f(r)g^2(r) & \approx &\lp 1+\frac{r^2}{l^2}+\frac{q^2}{3r^4}-\frac{m}{r^2} \rp \ls 1-c_1\frac{2 r^2+l^2}{3l^4\sqrt{3(r^4+l^2 r^2)}} \rs^2\nn\\&\approx& \lp 1-\frac{2 c_1}{3 \sqrt{3} l^4}\rp^2+\frac{\lp 1-\frac{2 c_1}{3 \sqrt{3} l^4}\rp^2 r^2}{l^2}+\lp 2 \sqrt{3} c_1-9 l^4\rp\times\nn\\&&\frac{ \ls\sqrt{3} c_1 \lp l^2-4 m\rp+18 l^4 m\rs}{162 l^8 r^2}+\frac{\lp q-\frac{2 c_1 q}{3 \sqrt{3} l^4}\rp^2}{3 r^4}+\mathcal{O}\lp r^{-6} \rp .\label{sgttmius}
\eea
The infinite redshift surfaces $ r=r_{s_{i}} $ are defined by roots of $ g_{tt}=0 $. It can be seen from the first line of Eq. (\ref{sgttmius}) that for $ c_1 \leq 0 $, the infinite redshift surfaces coincide with the horizons \ie $ r_{s_{i}}=r_\pm $. On the other side, for $ c_1 > 0 $, the function $ g_{tt} $ admits an additional root $ r_{s_{3}} =\frac{1}{\sqrt{2}} \sqrt{3 l^{12}/\sqrt{9 l^{20}-4 c_1^2 l^{12}/3}-l^2} $, as illustrated in Fig. \ref{fig3}. When $ r\rightarrow 0 $, both Ricci and Kretschmann scalars diverge, while in the limit $ r\rightarrow \infty $ they go to the values: $ R=-20/l^2 $, and $ R_{\m \n \r \s}R^{\m \n \r \s}=40/l^2 $, respectively. Therefore, there is a curvature singularity at $ r = 0 $. 
The electric field at large $ r $ is also given by
\be
E(r)=-F_{tr}\approx \frac{q}{r^3}\ls  1-c_1\frac{2r^2+l^2}{3l^{4}\sqrt{3(r^4+l^2 r^2)}} \rs.
\ee
The diagram of $ E(r) $ is plotted in Fig. \ref{fig4}. As one can see, the electric field diverges for small $ r $, and tends to zero at large $ r $ limit. 

In the following, we explore the uncharged black hole solutions. To do so, we set $ q=0 $ in the field equations. In this case, we observe that the gravitational field equations (\ref{stteq})-(\ref{spspsththphpheq}) have the following solutions 
\be
f(r)=1+\frac{r^2}{l^2}-\frac{m}{r^2},\label{usolf}
\ee
\be
g(r)=1+c_2\frac{2r^2+l^2}{\sqrt{\lvert r^4+l^2(r^2-m)\rvert}}.\label{usolg}
\ee  
The integration constant $ c_2 $ in the above equation, imposes the effect of mimetic field in the solution. The horizon is determined by real roots of $ g^{rr}=f(r_+)=0 $ which is given by $ r_+=\frac{1}{\sqrt{2}}\lp l \sqrt{\sqrt{l^2+4m}-1} \rp $. The curvature invariants have finite values at $ r_+ $ and as a result this is a coordinate singularity, not a curvature one. The sign of $ \lvert r^4+l^2(r^2-m)\rvert $ in the denominator of second term of Eq. (\ref{usolg}) depends on whether one consider $ r < r_+ $ or  $ r > r_+ $ regions. 
By expanding $ g(r) $ for large $ r $, one arrives at
\be
g(r)\approx 1+2 c_2+\frac{c_2 l^2\lp l^2+ 4 m\rp}{4r^4}+\mathcal{O}\lp r^{-6}\rp.
\ee
When $ r\rightarrow \infty $, $ g(r)\rightarrow 1+2c_2 $, that indicates the mimetic field incorporates in the metric function $ g(r) $ through the constant $ c_2 $. Also, the $ tt $ component of metric \ie the function $ f(r)g^2(r) $ reads
\be
f(r)g^2(r)=\frac{\ls c_2\lp l^2+2 r^2\rp+\sqrt{\lvert r^4+l^2 \lp r^2-m\rp\rvert}\rs^2}{l^2 r^2},
\ee  
with the following expansion at large $ r $
\bea
f(r)g^2(r) & \approx &  \lp 1+2 c_2\rp^2+\frac{\lp 1+2 c_2\rp^2 }{l^2}r^2+\frac{\lp 1+2 c_2\rp \lp c_2 l^2-2 m\rp}{2 r^2}+\mathcal{O}\lp r^{-6}\rp.
\eea 
The infinite redshift surfaces are the roots of $ g_{tt}=0 $, they are
\be
r_{s_{i}}=\sqrt{\frac{\lp 1-4 c_2^2\rp l^2\pm\sqrt{\lp 1-4 c_2^2\rp l^2 \lp l^2+4 m\rp}}{8 c_2^2-2}},
\ee
where $ i=1,2 $. The Ricci scalar and the Kretschmann invariant are also given by the following expressions:
\be
R=\frac{2 c_2 \ls l^4 -20 r^4+2l^2\lp2m-5r^2\rp \rs-20 r^2 \sqrt{\lvert r^4+l^2 \lp r^2-m\rp \rvert}}{ l^2 r^2\ls c_2\lp l^2+2r^2\rp+\sqrt{\lvert r^4+l^2 \lp r^2-m\rp \rvert}\rs},
\ee
\bea
R_{\m \n \r \s}R^{\m \n \r \s}&=&\frac{1}{l^4 r^8 \ls c_2\lp l^2+2r^2\rp+\sqrt{\lvert r^4+l^2 \lp r^2-m\rp \rvert}\rs ^2}\times\nn\\&& \left\{ -8 \lp 5 r^8+9 l^4 m^2\rp \ls l^2 \lp m-r^2\rp-r^4\rs \right.\nn\\&&+4 c_2^2 \ls l^8 \lp-18 r^2 m+7 r^4+18 m^2\rp+l^6 \lp 20 r^4 m-2 r^6\rp\right.\nn\\&&\left. +r^4 \lp 40 r^4 m^2-8 r^6 m+6 r^8\rp+8 r^8 l^2 \lp 5 r^2-2 m \rp+40 r^{12}\rs\nn\\&&+8 c_2 \sqrt{\lvert r^4+l^2 \ls r^2-m\rp \rvert} \lp 9 l^6 m \lp 2 m-r^2\rp-r^6 l^4\right.\nn\\&&\left.\left.+l^2 \lp 10 r^8-4 r^6 m\rp+20 r^{10}\rs\right\}.
\eea 
In the limit $ r \rightarrow 0 $, both Ricci and Kretschmann invariants diverge. As $ r \rightarrow \infty $, they go to $ -20/l^2 $ and $ 40/l^4 $, respectively. Therefore, there is an essential singularity at $ r = 0 $.

\subsection{Charged rotating solutions in cylindrical symmetric spacetimes}

Now, let us investigate charged rotating solutions with cylindrical symmetric spacetimes. In doing so, we consider the following five-dimensional metric \cite{Lemos:1994xp,Lemos:1995cm,Awad:2002cz}
\bea
ds^2&=&-f(r)g^2(r)\lp \Xi dt -\sum_{i=1}^{2}a_i d\varphi_i \rp ^2+\frac{1}{f(r)}dr^2\nn\\&&+\frac{r^2}{l^4} \sum_{i=1}^{2} \lp a_i dt -\Xi l^2 d\varphi_i \rp ^2-\frac{r^2}{l^2}\sum_{i<j}^{2}\lp a_i d\varphi_j-a_j d\varphi_i\rp^2+\frac{r^2}{l^2}dz^2,\label{rotmetric}
\eea
where $ \Xi=\sqrt{1+\sum_{i=1}^{2}a_i^2/l^2} $. The constants $ a_i $ and $ l $ have dimensions of length and as we will see later, the former are rotation parameters and the latter can be interpreted as the AdS radius. Here also, as spherically symmetric metric (\ref{sssmetric}), we have considered a new function $ g(r) $ because of an extra gravitational degree of freedom due to the mimetic field. The ranges of the time and radial coordinates are $ -\infty<t<\infty,0\leq r<\infty  $. The topology of horizon can be \begin{enumerate}[(i)] \item the flat $ 3 $-torus $ T^3 $ with topology $ S^1\times S^1\times S^1 $ and the ranges $ 0\leq \varphi_1<2\pi,0\leq \varphi_2<2\pi,0\leq z<2\pi l $, \item the three-dimensional cylinder with topology $ \mathbb{R}\times S^1\times S^1 $ and the ranges $ 0\leq \varphi_1<2\pi,0\leq \varphi_2<2\pi,-\infty<z<\infty $, \item  the three-dimensional infinite plane with topology $ \mathbb{R}^3 $ and the ranges $ -\infty<\varphi_1<\infty,-\infty<\varphi_2<\infty,-\infty<z<\infty $.\end{enumerate}
Here, we consider the topologies (i) and (ii). The former describes uncharged rotating solutions, while the latter represents charged rotating solutions. The gauge potential is also given by
\be
A_{\m}=h(r)\lp \Xi \d_{\m}^{0}-  \sum_{i=1}^{2} a_i \d_{\m}^{\varphi_i} \rp.
\ee
Having had the above gauge potential, the nonvanishing components of electromagnetic field strength tensor take the following form 
\be
F_{tr}=-\Xi h'(r),F_{\varphi_1 r}=a_1 h'(r),F_{\varphi_2 r}=a_2 h'(r).\label{nonzeroF}
\ee 
By substituting the gauge potential into the Maxwell field equation (\ref{Feq}), one finds the $ t $, $ \ph_1 $ and $ \ph_2 $ components of Maxwell equation yield the same differential equation for the function $ h(r) $, that is  
\be
g\lp rh''+3 h'\rp-rg'h'=0,
\ee 
which admits the following solution 
\be  
h'(r)=\frac{q}{r^3} g(r),\label{hsol}
\ee 
where we replace the constant of integration by the electric charge $ q $. Plugging the solution (\ref{hsol}) into the nonzero components of Maxwell tensor (\ref{nonzeroF}) and inserting the results into Eq. (\ref{geqfinal}), one arrives at the following seven independent differential equations
\bea
&& r^7 l^2 \lp a_1^2+a_2^2\rp  \ls 3 r f' g'+2 f \lp r g''+2 g'\rp \rs \nn\\&&+r^2 \lp a_1^2+a_2^2\rp  g \lp r^6 l^2 f''+4 r^5 l^2 f'+2 r^4 l^2 f -12 r^6-2 q^2 l^2\rp \nn\\&&-\Xi ^2 l^4 f g^3 \lp 3 r^5 l^2 f'+6 r^4 l^2 f-12 r^6+2 q^2 l^2\rp=0\label{tteq},
\eea
\bea
&& -r^7 l^2 \ls \lp 3 r f'+4 f\rp  g'+2 r f g''\rs +l^2 f g^3 \ls 3 r^4 l^2 \lp r f'+2 f\rp +2 \lp q^2 l^2-6 r^6\rp \rs \nn\\&&+g \{ 2 r^2 \lp 6 r^6+q^2 l^2\rp -r^6 l^2 \ls r \lp r f''+4 f'\rp +2 f\rs \}=0\label{rreq},
\eea
\bea
g \{ -3 r^4 l^2 \ls r \lp r f''+5 f'\rp +4 f\rs +48 r^6+4 q^2 l^2\} -3 r^5 l^2 \ls \lp 3 r f'+4 f\rp  g'+2 r f g''\rs=0 \label{phi1tandphi2teq},
\eea
\bea
&&a_1^2 l^2 f g^3 \ls 3 r^4 l^2 \lp r f'+2 f\rp +2 \lp q^2 l^2-6 r^6\rp \rs \nn\\&&+r^7 l^2 (a_2+\Xi  l) (a_2-\Xi  l) \ls \lp 3 r f'+4 f\rp  g'+2 r f g''\rs \nn\\&&+r^2 g (a_2+\Xi  l) (a_2-\Xi  l) \{ r^4 l^2 \ls r \lp r f''+4 f'\rp +2 f\rs -2 \lp 6 r^6+q^2 l^2\rp \}=0\label{phi1phi1eq},
\eea
\bea
&&r^7 l^2 \ls \lp 3 r f'+4 f\rp  g'+2 r f g''\rs -l^2 f g^3 \ls 3 r^4 l^2 \lp r f'+2 f\rp +2 \lp q^2 l^2-6 r^6\rp \rs \nn\\&& +g \{ r^6 l^2 \ls r \lp r f''+4 f'\rp +2 f\rs -2 \lp q^2 r^2 l^2+6 r^8\rp \}=0 \label{phi1phi2eq},
\eea
\bea
&&a_2^2 l^2 f g^3 \ls 3 r^4 l^2 \lp r f'+2 f\rp +2 \lp q^2 l^2-6 r^6\rp \rs \nn\\&&+r^7 l^2 (a_1+\Xi  l) (a_1-\Xi  l) \ls \lp 3 r f'+4 f\rp  g'+2 r f g''\rs \nn\\&&+r^2 g (a_1+\Xi  l) (a_1-\Xi  l) \{ r^4 l^2 \ls r \lp r f''+4 f'\rp +2 f\rs -2 \lp 6 r^6+q^2 l^2\rp \} =0\label{phi2phi2eq},
\eea
\be
r^7 l^2 \ls \lp 3 r f'+4 f\rp  g'+2 r f g''\rs +g \{ r^6 l^2 \ls r \lp r f''+4 f'\rp +2 f\rs -2 \lp q^2 r^2 l^2+6 r^8\rp\}=0 \label{zzeq}.
\ee
The next step is to solve the above field equations and obtain the unknown functions $ f(r) $ and $ g(r) $. To do so, we subtract Eq. (\ref{zzeq}) from Eq. (\ref{phi1phi2eq}). Doing so, yields the following equation on $ f(r) $
\bea
-l^2 f g^3 \ls 3 r^4 l^2 \lp r f'+2 f\rp +2 \lp q^2 l^2-6 r^6\rp \rs=0.\label{chgeq}
\eea
Solving the equation leads to 
\be
f(r)=\frac{r^2}{l^2}-\frac{m}{r^2}+\frac{q^2}{3r^4},\label{fsol}
\ee
where the integration constant $ m $ is the mass of solutions.
\begin{figure}[htp]
	\begin{center}
		\includegraphics[width=10cm]{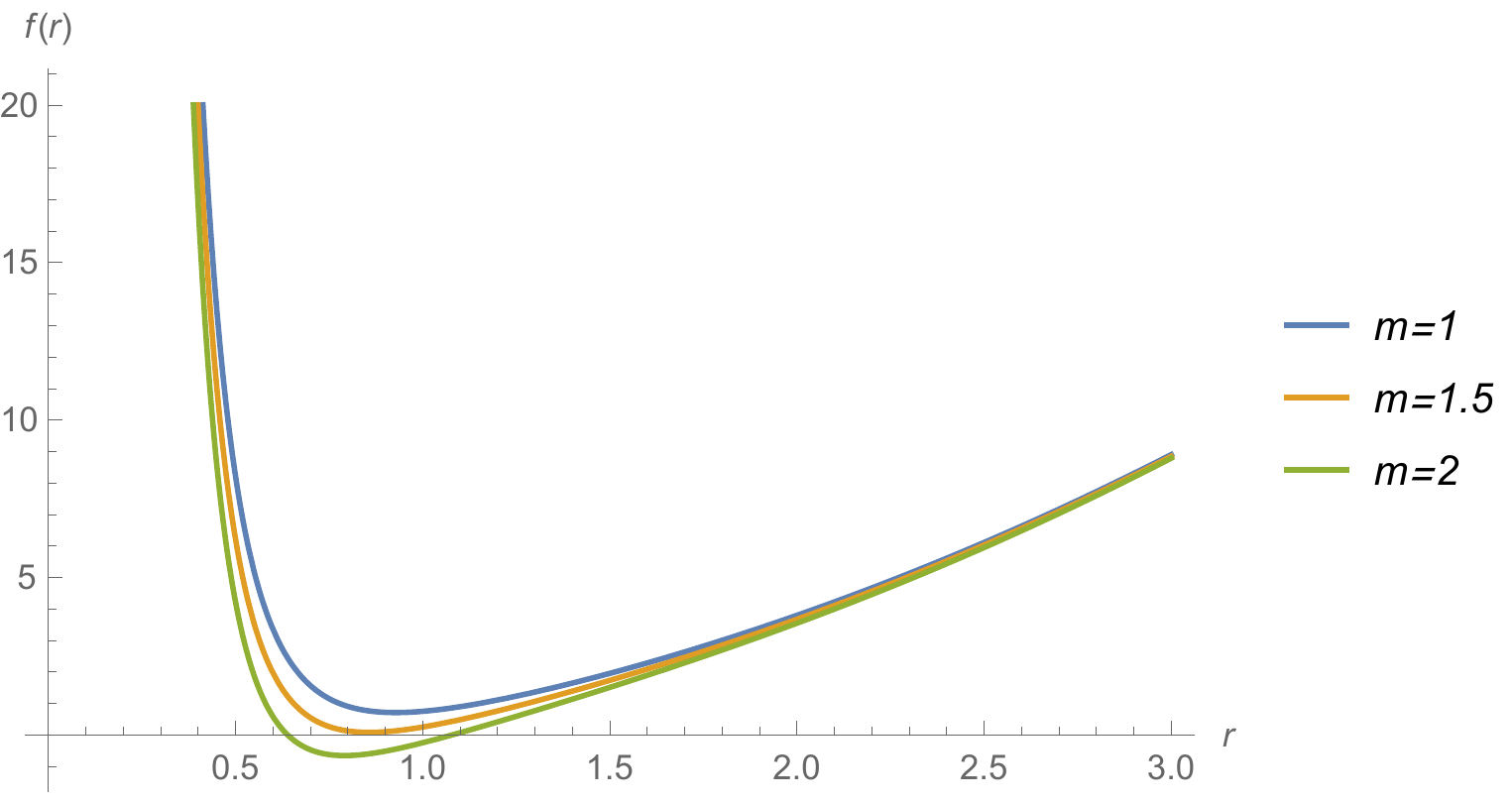}
		\caption{The diagram of function $f(r)$ for five-dimensional charged rotating solutions in mimetic gravity with $l=1$, $ q=1.5 $ and different values of $ m $.}\label{fig5}
	\end{center}
\end{figure}

Substituting the function $ f(r) $ from Eq. (\ref{fsol}) into Eq. (\ref{rreq}), leads to the following equation for the metric function $ g(r) $
\be
\ls l^2 \lp -3 m r^2+4 q^2\rp-15 r^6\rs g'+\ls r l^2 \lp -3 m r^2+q^2\rp+3 r^7 \rs g''=0,
\ee
with the solution
\be
g(r)=1+c_3\int \frac{r^4 dr}{\lvert 3r^6+l^2\lp q^2-3 m r^2\rp \rvert^{3/2}},\label{gsol1}
\ee
where, as already mentioned, the constant $ c_4 $ reflects the imprint of mimetic field in the solutions. It also easy to show that the solutions (\ref{fsol}) and (\ref{gsol1}) satisfy Eq. (\ref{phieqfinal}) in five dimensions. For large $ r $, the absolute value in the integrand of (\ref{gsol1}) can be approximated by $ 3r^6 $. Therefore, we can rewrite the function $ g(r) $ as
\be
g(r)\approx 1-\frac{c_3}{12\sqrt{3}r^4}.
\ee 
\begin{figure}[htp]
	\begin{center}
		\includegraphics[width=10cm]{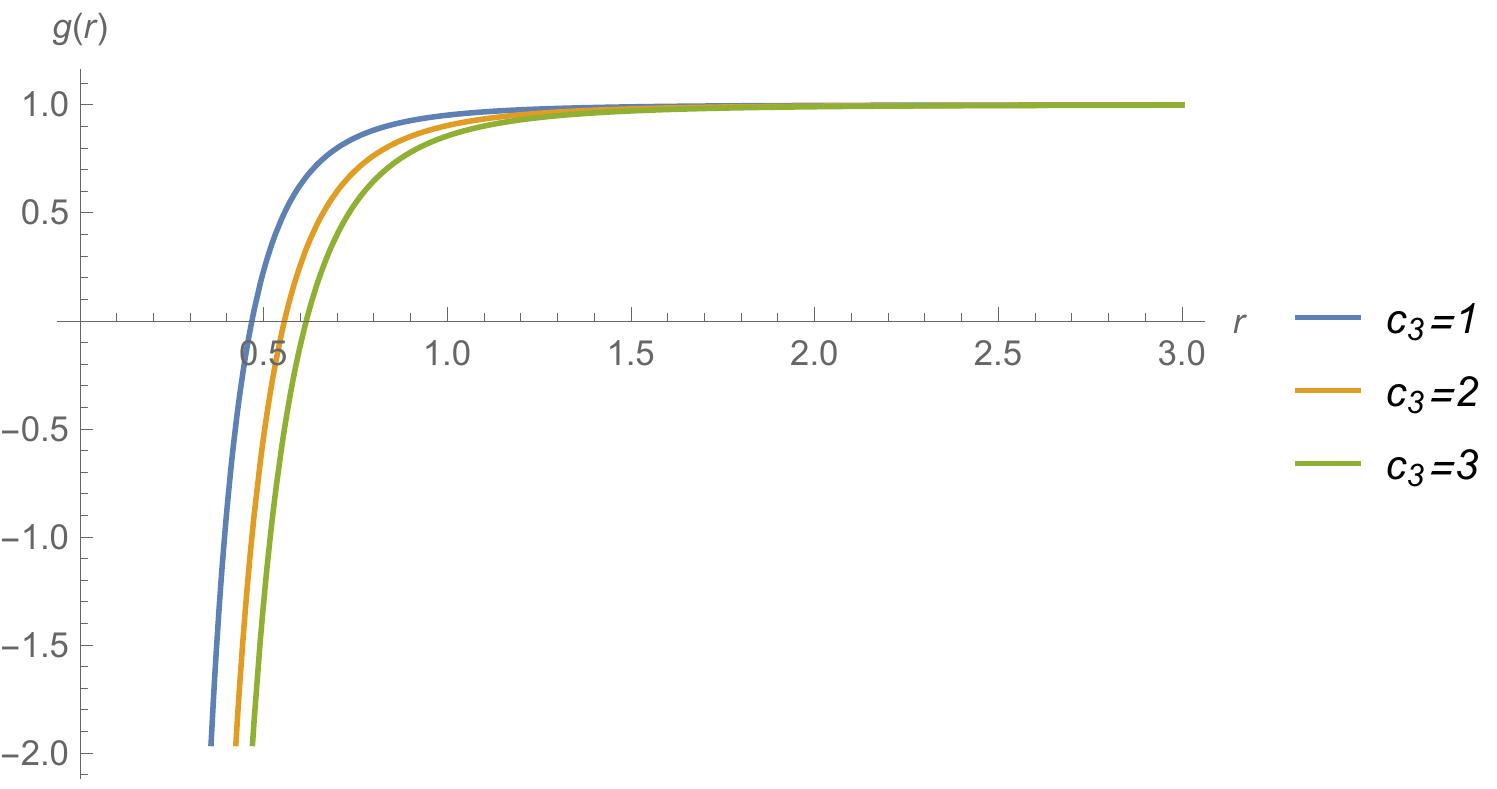}
		\caption{The diagram of function $g(r)$ for five-dimensional charged rotating solutions in mimetic gravity and different values of $ c_3 $.}\label{fig6}
	\end{center}
\end{figure}

In this limit the metric Function $ f(r) g^2(r)$ is also given by 
\be
f(r) g^2(r)\approx\frac{r^2}{l^2}-\frac{m}{r^2}-\frac{c_3}{6\sqrt{3}l^2r^2}+\frac{q^2}{3r^4}+\mathcal{O}\lp\frac{1}{r^6}\rp,
\ee
In the above solutions, the term containing $ c_3 $ is the impact of mimetic field on charged rotating solution of Einstein gravity in five dimensions.
\begin{figure}[htp]
	\begin{center}
		\includegraphics[width=10cm]{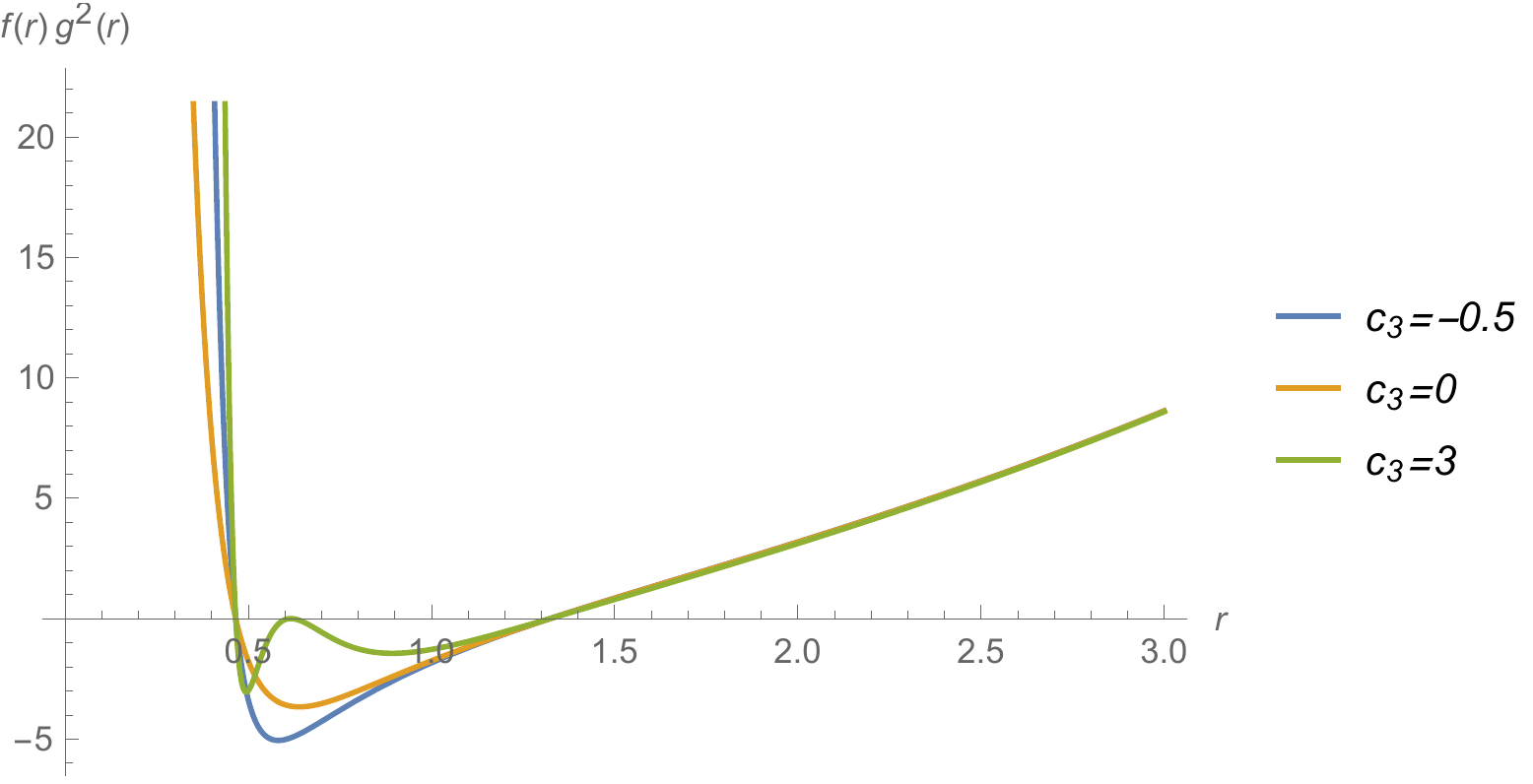}
		\caption{The behavior of function $f(r)g^2(r)$ for five-dimensional charged rotating solutions in mimetic gravity and different values of $ c_3 $. Here we choose $ m=3.5,l=1,q=1.5 $.}\label{fig7}
	\end{center}
\end{figure}

On the other side, when $ r\rightarrow 0 $, one can neglect the term $ 3r^6 $ in the absolute value term in the integrand of Eq. (\ref{gsol1}) and therefore the function $ g(r) $, in this limit, is given by
\be
g(r)\approx 1+ \frac{c_3 r^5}{5 l^3 q^3}+\cdots,
\ee
and hence
\be
f(r) g^2(r) \approx \frac{q^2}{3 r^4}-\frac{m}{r^2} +\frac{2 c_3 r}{15 q l^3}+\frac{r^2}{l^2}-\frac{2 c_3 r^3 m}{5 q^3 l^3}+\cdots.
\ee
It can be seen that, near the singularity $ r\rightarrow 0 $, metric function $ f(r) g^2(r) $ diverges, as was expected.

\begin{figure}[htp]
	\begin{center}
		\includegraphics[width=10cm]{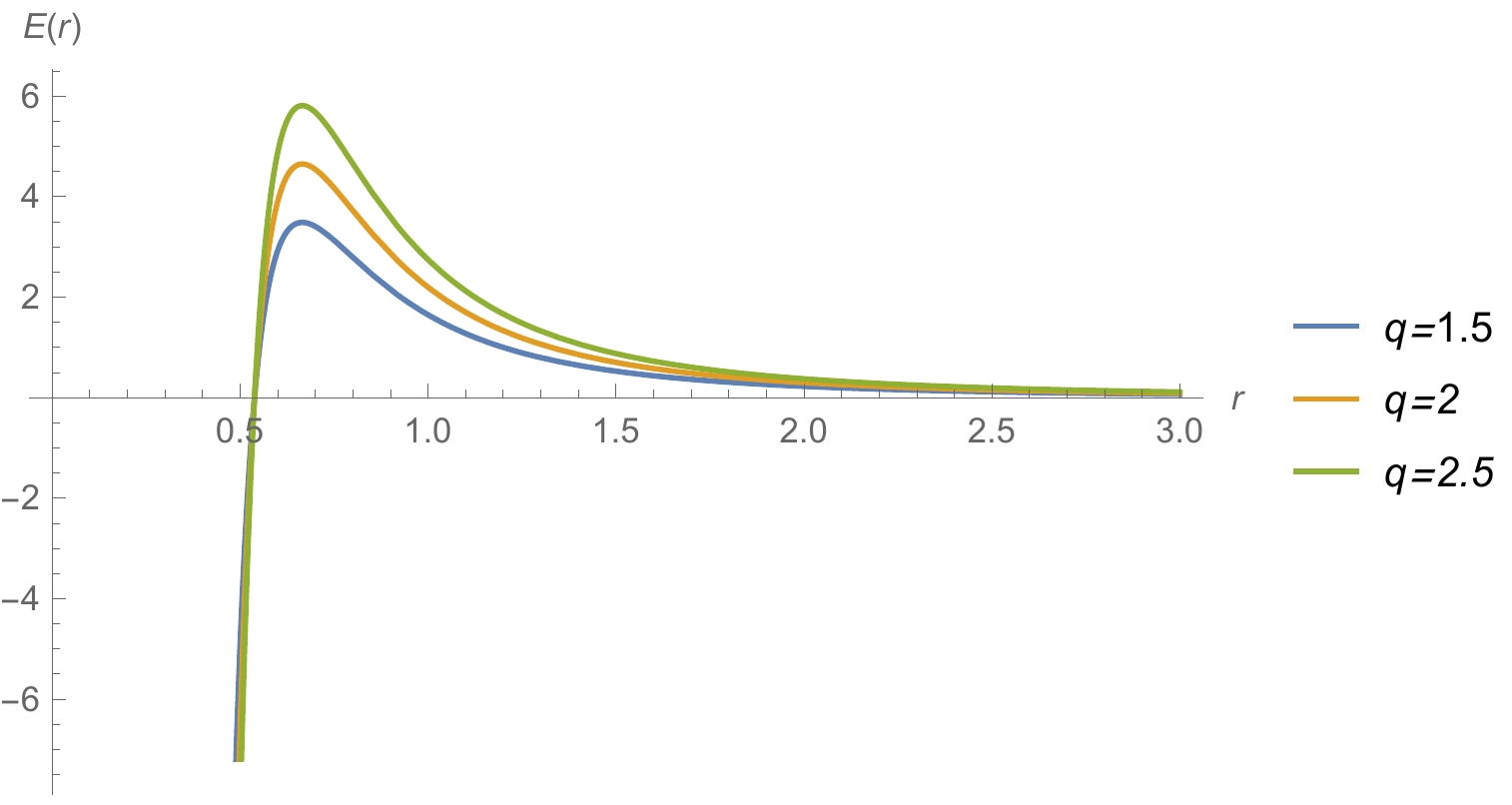}
		\caption{The behavior of Electric field  $E(r)$ for five-dimensional charged rotating solutions in mimetic gravity with $ c_3=1.75 $, $ \Xi=1.2 $ and different values of $ q $.}\label{fig8}
	\end{center}
\end{figure}
\begin{figure}[htp]
	\begin{center}
		\includegraphics[width=10cm]{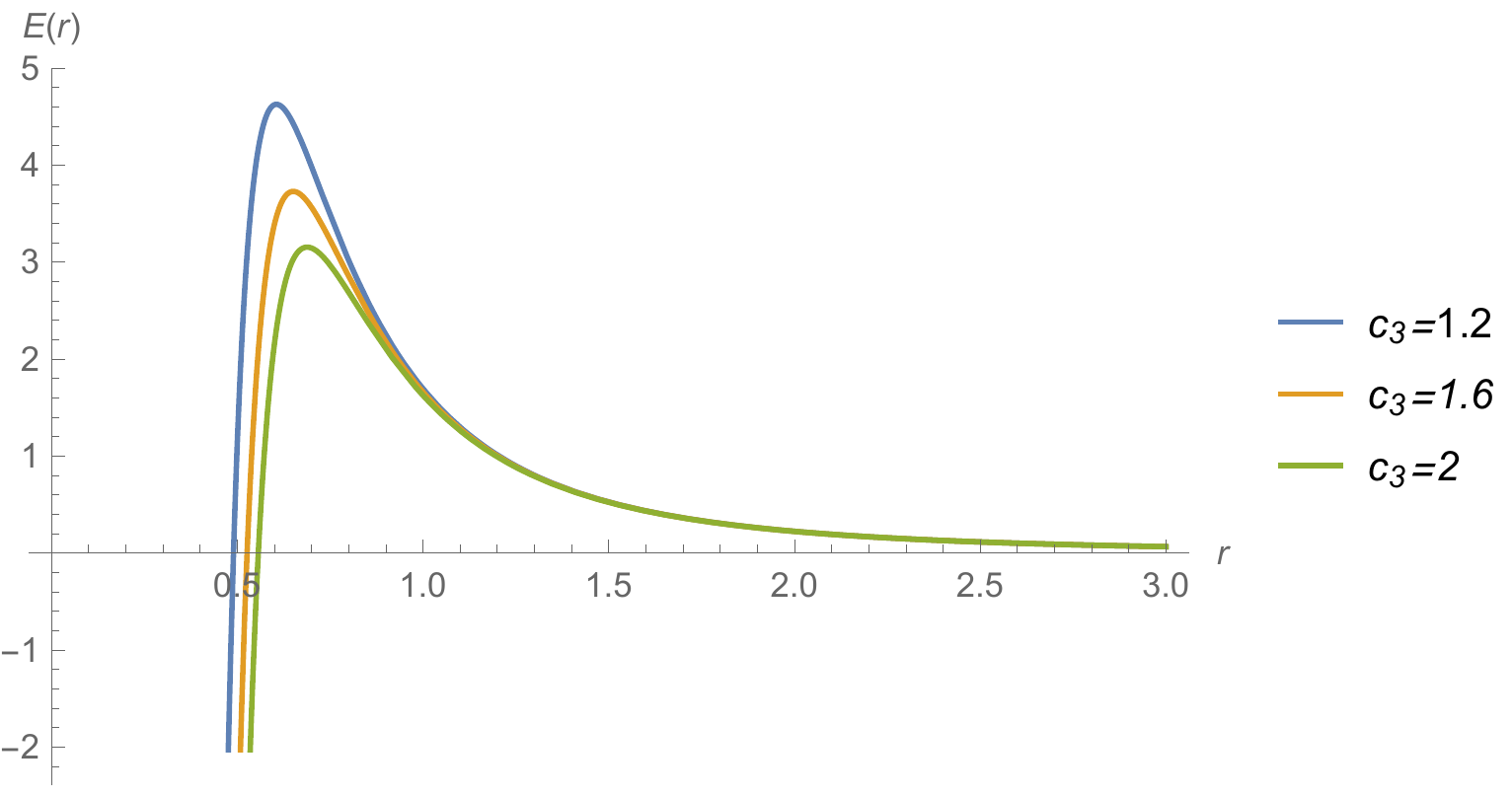}
		\caption{The behavior of Electric field  $E(r)$ for five-dimensional charged rotating solutions in mimetic gravity with $ q=1.5 $, $ \Xi=1.2 $ and different values of $ c_3 $.}\label{fig9}
	\end{center}
\end{figure}

The horizon is the real root of equation $ f(r)=0 $, for which the metric function $ f(r) $ is given by (\ref{fsol}). Depending on the different values of parameters, this equation may have zero, one or two roots. The cases with zero or one root correspond to naked singularity and extremal solutions, respectively. When this equation has two roots, we encounter a solution with two inner and outer horizon. For $ r\rightarrow 0 $, both Ricci and Kretschmann scalars diverge while in the limit $ r\rightarrow \infty $, they goes to the values: $ R=-20/l^2 $ and $ R_{\m \n \r \s}R^{\m \n \r \s}=40/l^2 $. Therefore, there is a curvature singularity at $ r = 0 $. The electric field at large $ r $ is also given by
\be
E(r)=-F_{tr}\approx \frac{q\Xi}{r^3}\lp 1-\frac{c_3}{12\sqrt{3}r^4} \rp.
\ee
The behavior of $ E(r) $ is plotted in Figs. \ref{fig8} and \ref{fig9}. As can be seen, the electric field diverges for small $ r $ and tends to zero at large $ r $ limit. 

The uncharged solutions can be found by setting $ q=0 $ in Eq. (\ref{chgeq}),
\be
\lp -3 m l^2 r^2-15 r^6\rp g'+\lp -3 m l^2 r^3 +3 r^7 \rp g''=0,\label{unchgeq}
\ee
with the following solution 
\be
g(r)=1+ c_4 \frac{ r^2}{\sqrt{\lvert r^4-m l^2 \rvert}}.\label{gsol2}
\ee
To investigate the asymptotic behavior of solutions, we consider the nonrotating case ($ a_1 = a_2 = 0 $), in the above solutions. The $ tt $ component of metric is also given by
\bea
f(r)g^2(r) = \frac{\lp c_4 r^2+\sqrt{\lvert r^4-m l^2 \rvert}\rp^2}{l^2 r^2}.
\eea
Expanding $ g(r) $ and $ f(r)g^2(r) $ for large values of $ r $ leads to
\be
g(r) \approx 1+c_4+\frac{ c_4 m l^2 }{2 r^4}+\mathcal{O}\lp r^{-8}\rp,
\ee
\be
f(r)g^2(r) \approx \frac{\lp 1+c_4 \rp^2}{l^2} r^2-\frac{\lp 1+c_4 \rp m}{r^2} +\mathcal{O}\lp r^{-6}\rp.
\ee
The Ricci scalar and the Kretschmann invariant take the following form
\be
R=\frac{4 \lp c_4 l^2 m-5  c_4 r^4-5 r^2 \sqrt{\lvert r^4-m l^2 \rvert}\rp}{r^2 l^2 \lp c_4 r^2+\sqrt{\lvert r^4-m l^2 \rvert}\rp},
\ee
\bea
R_{\m \n \r \s}R^{\m \n \r \s}&=& \frac{8}{l^4 r^8 \ls\lp c_4^2-1\rp r^4+l^2 m\rs^2} \times \\&& \left \{-2 \lp c_4^2-1\rp r^{10} l^2 m \ls\lp c_4^2-5\rp r^2- c_4 \sqrt{\lvert r^4-m l^2 \rvert}\rs \right. \nn\\&& \left.-2 l^6 m^3 r^2 \ls\lp 7  c_4^2+9\rp r^2-9  c_4 \sqrt{\lvert r^4-m l^2 \rvert}\rs+5 \lp c_4^2-1\rp^2 r^{16}\right. \nn\\&& \left.+ l^4 m^2 r^6\ls\lp5  c_4^4+12  c_4^2+14\rp r^2-2  c_4 \lp 5  c_4^2+8\rp \sqrt{\lvert r^4-m l^2 \rvert}\rs+9 l^8 m^4\right \}\nn.
\eea
In the limit $ r \rightarrow 0 $, both Ricci and Kretschmann invariants diverge. As $ r \rightarrow \infty $, they go to $ -20/l^2 $ and $ 40/l^4 $, respectively. This confirms that $ r = 0 $ is an essential singularity of spacetime.

In the following, we are going to find some physical quantities of solutions such as temperature, entropy, mass and angular momenta. By setting $ t\rightarrow i \t $, one can obtain the Euclideanized version of the metric (\ref{sssmetric}) which implies that the period of Euclidean time to be as $ \t \sim \t + \b_+ $. Thus, the inverse Hawking temperature takes the following form \cite{Sheykhi:2020fqf}
\be
\b_+=\frac{1}{T}=\frac{4 \pi \Xi}{g(r_+)f'(r_+)},
\ee
where $ r_+ $ is the largest root of $ f(r) $. Upon Euclidean continuation, we must take $ a_j \rightarrow i a_j $, as well. The periodicity of Euclideanized time $ \t \sim \t + \b_+ $ also leads to the periodicity of angular coordinates as $ \ph_j \sim \ph_j + i\b_+ {\Omega_+}_j $, where
\be
{\Omega_+}_j=\frac{a_j}{\Xi l^2},
\ee
are the angular velocities at the horizon \cite{Hawking:1998kw}. The above result for inverse Hawking temperature can also be obtained by computing the surface gravity via the following formula:
\be
\b_+=\frac{2\pi}{\k}=\frac{2\pi}{\sqrt{-\frac{1}{2}(\nabla^\m \chi^\n)(\nabla_\m \chi_\n)}},
\ee
where the Killing vector is given by $ \chi^\n=(1,0,{\Omega_+}_1,{\Omega_+}_2,0) $  associated with the Killing horizon $ {\cal S}(r=r_+) $. The entropy per unit volume of horizon for metric (\ref{sssmetric}) is given by the quarter of horizon area
\be
S=\frac{A_h}{4}=\frac{r_+^3 \Xi}{4l}.
\ee
To obtain thermodynamical quantities, we should add the Gibbons–Hawking boundary term to $ 5 $-dimensional version of action (\ref{Daction}) and then get Einstein-Maxwell equations by varying the bulk metric with fixed boundary metric. Here, the boundary action is given by 
\be
I_s=\frac{1}{8\pi}\int_{\pa \cal{M}} d^4x \sqrt{-\g} \Th,
\ee
where $ \g_{ab} $ is the boundary metric and $ \Th $ is the trace of the extrinsic curvature $ \Th_{ab} $ of the boundary. We use the counterterm approach \cite{Henningson:1998gx,Balasubramanian:1999re} to eliminate the divergences of action. In this approach we add some local surface integrals to the action to make it finite. The following counterterm get a finite action up to seven dimensions \cite{Awad:2002cz}
\be
I_{ct}=\frac{1}{8\pi}\int_{\pa \cal{M}} d^4x \sqrt{-\g} \lp \frac{3}{l}-\frac{l}{4} \cal{R}+\cdots\rp,\label{5ctaction}
\ee
where dots denote the terms which are needed in higher dimensions. Here, $ \cal{R} $ and $ {\cal R}_{ab} $ are the Ricci scalar and tensor for the boundary metric $ \g $. By varying the finite action with respect to boundary metric $ \g_{\m\n} $, we can find a divergence-free stress tensor which has been introduced by Brown and York \cite{Brown:1992br}
\be
T^{ab}=\frac{1}{8\pi}\lp \Th^{ab}-\Th \g^{ab}+\frac{3}{l}\g^{ab}-\frac{l}{2} {\cal G}^{ab}+\cdots\rp,\label{5enmomtens}
\ee
where $ {\cal G}_{ab}={\cal R}_{ab}-{\cal R}\g_{ab}/2 $ is the Einstein tensor of boundary metric $ \g_{ab} $. The above stress tensor is divergence-free for dimensions less than six, but one can add more counterterms to get a finite action in higher dimensions \cite{Kraus:1999di}. For asymptotically AdS solutions with flat horizons, the only nonvanishing counterterm is the first term in (\ref{5ctaction}) which yields the stress-energy tensor (\ref{5enmomtens}) up to the third term. Using this stress tensor, one can define the quasilocal conserved quantities for an asymptotically AdS spacetime. The conserved charge associated to a Killing vector $ \xi_a $ is given by
\be
Q_\xi=\int_{\S} d^3x\sqrt{\s} u^a T_{ab}\xi^b,\label{consch}
\ee
where $ u_a =-N\d^{0}_{a} $, while $ N $ and $ \s $ are the lapse function and the spacelike metric which appear in the ADM–like decomposition of the boundary metric
\be
\g_{ij}dx^i dx^j=-N^2 dt^2+\s_{ab}\lp dx^a + V^a dt\rp \lp dx^b + V^b dt \rp.
\ee
Here, $ V^a $ is the shift vector. To obtain the total mass (energy) we should set $ \xi=\pa_t $ \ie the Killing vector conjugate to time coordinate $ t $ and to obtain the i{\it th} component of angular momentum we should set $ \xi=\pa_{\ph_i} $ \ie the Killing vector conjugate to angular coordinate $ \ph_i $. Using the definition (\ref{consch}) for conserved charges, we find the total mass per unit volume of horizon for charged solutions to be given by
\be
M=\frac{1}{16 \pi l}\ls \lp 4\Xi^2-1\rp m+\frac{2}{3\sqrt{3}l^2}(\Xi^2-1)c_3\rs,
\ee
while the angular momenta per unit volume of horizon are given by
\be
J_i=\frac{1}{16 \pi l}\lp 4 m  +\frac{2}{3\sqrt{3}l^2}c_3\rp \Xi a_i.
\ee
In the case of uncharged solutions, the total mass and angular momenta per unit volume of horizon are as follows
\be
M=\frac{1}{16 \pi l}\lp 4 \Xi^2-1+3 c_4\rp m,
\ee
\be
J_i=\frac{1}{4 \pi l} \Xi a_i m.
\ee

Finally, to find the electric charge of solutions, we determine the electric field by considering the projections of the electromagnetic tensor on a special hypersurface. The normal vectors to such a hypersurface are
\be
u^0 = \frac{1}{N} ,u^r = 0, u^i=-\frac{V^i}{N},
\ee
where $ N $ and $ V^i $ are the lapse function and shift vector, respectively. The electric field is also given by $ E^\m = g^{\m\r}F_{\r\n}u^{\n} $. The electric charge per unit volume of black string can be found by calculating the flux of the electric field at infinity, \ie
\bea
Q=\frac{q \Xi}{l}.
\eea 
The electric potential $ \Phi $, measured at infinity with respect to the horizon, is defined by \cite{Dehghani:2002rr}
\be
\Phi=A_a \chi^a\lvert_{r\rightarrow\infty}-A_a\chi^a\lvert_{r=r_+},
\ee
Calculating the above expression leads to
\bea
\Phi=\frac{q}{2 \Xi r_+^2}\lp1-\frac{\sqrt{3} c_3}{108 r_+^4}\rp. 
\eea

\section{Extension to higher dimensions}\label{ext}

Now, we are in a position to extend our solutions to $ D>4 $ dimensions. We first consider the spherically symmetric spacetimes for which the line element is given by the $ D $-dimensional version of metric (\ref{sssmetric}), where $ d\Omega^{2} $ is the line element of a ($ D-2 $)-sphere. The topology of event horizon and spacetime are therefore $ S^{(D-2)} $ and $ \mathbb{R}^2\times S^{(D-2)} $, respectively. The next step is to solve Eqs. (\ref{geqfinal}) and (\ref{phieqfinal}) for the given metric. To do so, we solve them for six and seven dimensions\footnote{The detail of calculations are omitted here for the sake of brevity.} and find the following structure for the metric functions $ f(r) $ and $ g(r) $ in $ D $ dimensions   
\be
f(r)=1+\frac{r^2}{l^2}-\frac{m}{r^{D-3}}+\frac{2q^2}{(D-3)(D-2)r^{2(D-3)}},\label{Dsssfsol}
\ee
\be
g(r)=1+c_1\int \frac{r^{2(D-3)} dr}{\left\lvert\frac{(D-3)(D-2)}{2}r^{2(D-2)}+l^2\ls q^2+\frac{(D-3)(D-2)}{2}r^{D-3}\lp r^{D-3}-m\rp\rs\right\rvert^{3/2}}.\label{Dsssgsol}
\ee
To obtain an analytical solution for $ g(r) $, one can examine the asymptotic behavior of function at large $ r $, that is
\be
g(r)\approx 1-c_1\frac{2 r \sqrt{\frac{2 r^2}{l^2}+2} \, _2F_1\lp \frac{3}{2},2-\frac{D}{2};3-\frac{D}{2};-\frac{r^2}{l^2}\rp}{(D-4) (D-3) (D-2) l^2 \sqrt{(D-3) (D-2) r^{2(D-3)} \lp r^2+l^2\rp}},\label{lDsssgsol}
\ee
where $ _2F_1(a,b;c;z) $ is the Gaussian hypergeometric function. In the case of uncharged solutions we get
\be
f(r)=1+\frac{r^2}{l^2}-\frac{m}{r^{D-3}},\label{unchDsfsol}
\ee
\be
g(r)=1+c_1\int \frac{r^{(D-3)/2} dr}{\left\lvert r^{(D-1)}+l^2 \lp r^{D-3}-m\rp\right\rvert^{3/2}}.\label{unchDsssgsol}
\ee
Here also the function $ g(r) $ takes the following form at large $ r $
\be
g(r)\approx 1-c_2\frac{r^{\frac{5-D}{2}} \, _2F_1\lp 1,\frac{3-D}{2};3-\frac{D}{2};-\frac{r^2}{l^2}\rp}{(D-4) l^2 \sqrt{r^{D-3} \lp r^2+l^2\rp}}.
\ee
We have found that for the charged and uncharged cases, in the limit $ r \rightarrow 0 $, both Ricci and Kretschmann invariants diverge, while as $ r \rightarrow \infty $ they go to $ -D(D-1)/l^2 $, and $ 2D(D-1)/l^4 $, respectively. As a result, $ r = 0 $ is an essential singularity of spacetime.

The solutions for cylindrically symmetric spacetimes can also be found by considering the line element (\ref{rotmetric}) in $ D $ dimensions. Here, the metric (\ref{rotmetric}) is modified by considering $ n $ rotation parameters rather than 2, and the relation between the number of rotation parameters and dimension is given by $ n=[(D-1)/2] $, where the bracket denotes the integer part. In this case, the possible topologies of horizon are given by, \begin{enumerate}[(i)] \item the flat $ n $-torus $ T^n $ with topology $ S^1\times \cdots\times S^1 $ and the ranges $ 0\leq \varphi_i<2\pi,0\leq z<2\pi l $, \item the $ (D-2) $-dimensional cylinder with topology $ \mathbb{R}\times S^1\times \cdots \times S^1 $ and the ranges $ 0\leq \varphi_i<2\pi,-\infty<z<\infty $, \item  the infinite plane $ \mathbb{R}^n $ with the ranges $ -\infty<\varphi_i<\infty,-\infty<z<\infty $.\end{enumerate}
where $ i=1,\cdots,n $. Here, as five-dimensional case, we consider the topology (i) and (ii). Here also we solve the Einstein-Maxwell field equations in the presence of mimetic field for six and seven dimensions and find the following $ D $-dimensional form for the metric functions $ f(r) $ and $ g(r) $    
\be
f(r)=\frac{r^2}{l^2}-\frac{m}{r^{D-3}}+\frac{2q^2}{(D-3)(D-2)r^{2(D-3)}},\label{Dfsol1}
\ee
\be
g(r)=1+c_3\int \frac{r^{2(D-3)} dr}{\left\lvert\frac{(D-3)(D-2)}{2}r^{2(D-2)}+l^2\ls q^2-\frac{(D-3)(D-2)}{2} m r^{D-3}\rs\right\rvert^{3/2}}.\label{Dgsol1}
\ee
At large $ r $, the function $ g(r) $ reads
\be
g(r)\approx 1-\ls \frac{2}{(D-3)(D-2)}\rs^{3/2}\frac{c_3}{D-1}\frac{1}{r^{D-1}}.
\ee
In the absence of Maxwell field, we get
\be
f(r)=\frac{r^2}{l^2}-\frac{m}{r^{D-3}},\label{unDfsol}
\ee
\be
g(r)=1+ c_4 \frac{ r^{(D-1)/2}}{\sqrt{\lvert r^{D-1}-m l^2 \rvert}},\label{ungsol2}
\ee
Here also, as spherically symmetric spacetimes, we observe that even in the absence of Maxwell field in the limit $ r \rightarrow 0 $, both Ricci and Kretschmann invariants diverge while as $ r \rightarrow \infty $, they go to $ -D(D-1)/l^2 $ and $ 2D(D-1)/l^4 $, respectively. This indicates that, there is a spacetime singularity at $ r=0 $. 

The entropy per unit volume of horizon is given by the quarter of horizon area \cite{Awad:2002cz}
\be
S=\frac{A_h}{4}=\frac{r_+^{D-2}\Xi}{4l}.
\ee
By redefinition of conserved charges in $ D $-dimensions, we find the total mass per unit volume of horizon for the charged solutions to be given by
\be
M=\frac{1}{16 \pi l}\left\{ \ls(D-1)\Xi^2-1\rs m+\frac{2}{l^2}\lp\frac{2}{(D-3)(D-2)}\rp^{3/2}\lp\Xi^2-1\rp c_3\right\},
\ee
while the angular momenta per unit volume of horizon are given by
\be
J_i=\frac{1}{16 \pi l}\ls(D-1) m +\frac{2}{l^2}\lp\frac{2}{(D-3)(D-2)}\rp^{3/2}c_3\rs \Xi a_i.
\ee
In the absence of Maxwell field the total mass and angular momenta per unit volume of horizon are given by 
\be
M=\frac{1}{16 \pi l}\ls(D-1) \Xi^2-1+(D-2)c_4\rs m,
\ee
\be
J_i=\frac{D-1}{16 \pi l} \Xi a_i m.
\ee
It can be seen that, total mass and angular momenta reduce to the corresponding ones in Einstein-Maxwell gravity in asymptotically AdS and cylindrically symmetric spacetime by setting $ c_3=c_4=0$ \cite{Awad:2002cz}. For $ D = 4 $ and in the absence of mimetic field, they reduce to the mass and angular momenta given in \cite{Lemos:1995cm,Dehghani:2002rr}. For $ D = 4 $, they also reduce to the mass and angular momenta in \cite{Sheykhi:2020fqf} by reparametrization of the constants $ c_3 $ and $ c_4 $.

\section{Concluding remarks}\label{con}

We have presented two class of charged static and rotating solutions of mimetic gravity in  $ D $ dimensions. In doing so, we have constructed the solutions in five, six and seven dimensions in spherically and cylindrically symmetric spacetimes and then extend them to higher dimensions. 

We observe that, for both class of solutions, the spacetime admits a curvature singularity in mimetic gravity. We also have found that both Ricci and Kretschmann invariants diverge at $ r = 0 $ and are finite at $ r \neq 0 $. Thus, we have an essential singularity at this point. 

We have used the conserved charges defined by Brown and York to calculate the total mass and angular momenta of these solutions. It also can be seen that, the mass and angular momenta for uncharged solutions can not be recovered from the charged ones by setting $ q=0 $. The reason is that, the function $ g(r) $ in the charged case is obtained with the approximation large $ r $, while in the uncharged case, we have not used any approximation in obtaining the function $ g(r) $. We also observe that, when $ D=4 $, our results for singularities, the asymptotic behaviour of Ricci and Kretschmann invariants, as well as mass and angular momenta are fully consistent with the corresponding ones in black string solutions obtained in \cite{Sheykhi:2020fqf}. 

For future works, one can consider the solutions in the presence of a variable potential. Finding the solutions with higher-derivative corrections to mimetic field \cite{Gorji:2017cai} is also an interesting feature. 

\appendix

\section*{Acknowledgements}\addcontentsline{toc}{section}{Acknowledgements}

We would like to thank the authors of Ref. \cite{Nutma:2013zea} for developing the excellent \emph{Mathematica} package ``xTras,'' which we have used extensively for symbolic calculations.


\providecommand{\href}[2]{#2}\begingroup\raggedright
\endgroup
\end{document}